\documentclass[journal=langd5, manuscript=article]{achemso}

\usepackage[version=3]{mhchem} 
\usepackage[T1]{fontenc}       
\usepackage{amssymb}
\usepackage[dvipsnames]{xcolor} 
\usepackage{braket}
\usepackage{multirow}
\usepackage{siunitx} 
\usepackage{adjustbox} 
\usepackage{xstring} 

\newcommand{\ie}{i.e.\ }
\newcommand{\eg}{e.g.\ }

\newcommand{\PM}{\text{PM}}
\newcommand{\MM}{\text{1MM}}
\newcommand{\MMM}{\text{2MM}}

\renewcommand{\vec}[1]{\mathbf{#1}} 
\newcommand{\vecs}[1]{\boldsymbol{#1}} 

\newcommand{\dod}[2]{\frac{\text{d} #1}{\text{d} #2}}

\newcommand{\edit}[1]{#1}
\newcommand{\fedit}[1]{#1}

\newcommand{\dep}{\text{dep}}
\newcommand{\undep}{0}
\newcommand{\targetwt}[1]{T_{wt,{#1}}}
\newcommand{\targetmut}[1]{T_{mut,{#1}}}

\definecolor{light-gray}{gray}{0.85} 
\newcommand{\wtriplet}[1]{\adjustbox{margin=0ex 0.25ex, 
bgcolor=light-gray}{#1}} 
\newcommand{\colormut}[1]{\StrMid{#1}{1}{1}\textcolor{red}{\textbf{\StrMid{#1}{2}{2}}}\StrMid{#1}{3}{3}}
\newcommand{\mtriplet}[1]{\wtriplet{\colormut{#1}}}

\author{Stefanos K.\ Nomidis}

\affiliation{Laboratory for Soft Matter and Biophysics, KU Leuven,  
Celestijnenlaan 200D, 3001 Leuven, Belgium}

\alsoaffiliation{Flemish Institute for Technological Research (VITO), Boeretang 
200, B-2400 Mol, Belgium}

\author{Michal Szymonik}

\affiliation{Flemish Institute for Technological Research (VITO), Boeretang 
200, B-2400 Mol, Belgium}

\author{Tom Venken}

\affiliation{Center for Cancer Biology, VIB, 3000 Leuven, Belgium} 

\alsoaffiliation{Laboratory of Translational Genetics, Department of Human 
Genetics, KU Leuven, 3000 Leuven, Belgium}

\author{Enrico Carlon}

\affiliation{Laboratory for Soft Matter and Biophysics, KU Leuven,  
Celestijnenlaan 200D, 3001 Leuven, Belgium}

\author{Jef Hooyberghs}

\affiliation{Flemish Institute for Technological Research (VITO), Boeretang 200, 
B-2400 Mol, Belgium}

\alsoaffiliation{Theoretical Physics, Hasselt University, Campus Diepenbeek, 
B-3590 Diepenbeek, Belgium}

\email{jef.hooyberghs@vito.be}

\title{Enhancing the performance of DNA surface-hybridization biosensors
through target depletion}

\begin{document}


 
 


\begin{abstract}
DNA surface-hybridization biosensors utilize the selective hybridization
of target sequences in solution to surface-immobilized probes.  In this
process, the target is usually assumed to be in excess, so that its
concentration does not significantly vary while hybridizing to the
surface-bound probes. If the target is initially at low concentrations
and/or if the number of probes is very large and have high affinity for
the target, the DNA in solution may get depleted. In this paper we analyze
the equilibrium and kinetics of hybridization of DNA biosensors in the
case of strong target depletion, by extending the Langmuir adsorption
model.  We focus, in particular, on the detection of a small amount of a
single-nucleotide ``mutant'' sequence (concentration $c_2$) in a solution, 
\edit{which differs by one or more nucleotides from}
an abundant ``wild-type'' sequence (concentration $c_1 \gg c_2$). We
show that depletion can give rise to a strongly-enhanced sensitivity
of the biosensors. Using representative values of rate constants and
hybridization free energies, we find that in the depletion regime one
could detect relative concentrations $c_2/c_1$ that are up to three orders
of magnitude smaller than in the conventional approach. The kinetics
is surprisingly rich, and exhibits a non-monotonic adsorption with no
counterpart in the no-depletion case. Finally, we show that, alongside
enhanced detection sensitivity, this approach offers the possibility of
sample enrichment, by substantially increasing the relative amount of
the mutant over the wild-type sequence.

\end{abstract}



\section{Introduction}

DNA hybridization, the binding of two single-stranded DNA molecules
to form a double-stranded helix, is a physico-chemical process of \edit{broad 
interest to disciplines ranging}
from fundamental to applied sciences and engineering. \edit{It is 
also central to many applications where detection or enrichment of 
specific target DNA molecules is required. E.g. in clinical diagnostics, 
which is typically targeted and not hypothesis-free, the detection of known 
DNA variants is of high importance~\cite{khod16}. These variants, e.g. 
DNA from a tumor, can sometimes differ in only a single nucleotide from 
the wild-type DNA of the healthy cells. For non-invasive tests from 
peripheral blood, devices must be specific enough to detect mutated 
DNA molecules in a background of wild-type DNA down to frequencies of 
0.1\% or less. This challenge drives new detection principles 
and enrichment strategies among which hybridization-based\cite{khod16}.} 
In these applications, single-stranded DNA probes are designed to bind to
the target molecules during a hybridization process. Often the probe
molecules are immobilized on a surface for detection purposes or for
further processing. Using the sequence-specific properties of the
process, specificity and sensitivity of the binding are two important
characteristics that can be aimed for. This is often challenging due
to the presence of cross-hybridization, which occurs when DNA molecules
resembling the sequence of the target molecules hybridize to the probes
and blur the detection or poison the enrichment.

Hybridization of targets to surface-immobilized probes can be physically
described by the Langmuir adsorption model, used extensively to
predict the equilibrium state of typical systems \cite{naef02, vain02,
held03, zhan03, halp04, bind05, burd06, carl06, nais09, hooy09, hadi12,
harr13}. In a standard Langmuir approach, the target concentration is
assumed to be constant, which is the case when it is large enough not to
be depleted due to hybridization with the probe molecules. In experimental
applications this assumption may be violated, and corrections need to be
applied to incorporate the reduced target concentration into the model.

\edit{This paper builds upon three previous works that considered such a
target-depletion effect on surface hybridization. Michel et al.\ and Ono et 
al.\ independently calculated the equilibrium intensity for the case of one 
target (Ono et al.\ also for two) hybridizing with a single 
probe~\cite{mich07, ono08}. Then, Burden and Binder performed a more systematic 
analysis, by distinguishing between local and global depletion, depending on 
whether depletion by a probe affects only itself or all other probes too, 
respectively~\cite{burd09}. The hybridization model by Michel et al.\ and Ono 
et al.\ fall under the former category. Our work assumes that hybridization is 
operating in a non diffusion limited regime and that depletion is global. 
For global depletion Burden and Binder presented a numerical scheme to calculate 
the equilibrium solution, by assuming the probe concentration to be identical among 
different probes.} 

\edit{Our work extends this result, by analytically deriving the equilibrium 
solution for an arbitrary number of targets and probes, under a realistic 
assumption (no probe saturation), and allowing for varying probe concentration. 
This allowed us to design an experimental setup that exploits target depletion, 
so as}
to enhance the performance of DNA biosensors. In particular, we focus
on typical situations interesting for diagnostic purposes, where the
sample to be analyzed contains a large amount of ``wild type'' sequence
at concentration $c_1$ and a much smaller amount of ``mutant'' sequence,
differing by a single nucleotide~\cite{ang12, moha17, su17, rapi17}. The
latter is at a concentration $c_2 \ll c_1$.  We discuss a minimal-design
strategy (Fig.~\ref{fig:setup}) and show how the depletion of the wild
type sequence may lead to an increased sensitivity \edit{(as confirmed by a 
practical demonstration)}, where the detection of the mutant becomes possible 
even for very small ratios $c_2/c_1$. \edit{We also show that this method can be 
utilized in order to achieve sample enrichment~\cite{zhan18}, by increasing the 
ratio of the captured mutant over the wild type target. Finally, we calculated 
both analytically and numerically the kinetics of the process and analyzed 
the rich resulting behavior.}


\section{Materials and methods}

In what follows, we will first review the standard Langmuir adsorption model, 
and then present a simple extension, which accounts for the depletion of the 
target sequence. Finally, we discuss how this problem can be analytically 
approached by introducing some useful approximations, without much loss of 
generality.

\subsubsection{Langmuir adsorption model}

The Langmuir adsorption model treats hybridization as a two-state
process. Among the several simplifications, such as the homogeneity of the
surface and the lack of interactions among adsorbates, the model assumes
that the concentration of the target sequences in solution is so large,
that it practically remains unchanged throughout the process. Let us
consider the simple case of one target type in solution, brought into
contact with a single probe type.  Denoting by $\theta$ the fraction of
hybridized probes, \ie the number of hybridized probes divided by the
total number of probes, the kinetics of the process is described by
\begin{equation}\label{eq:langmuir_ode_simple}
 \dod{\theta}{t} = k^+ (1 - \theta) c - k^- \theta,
\end{equation}
where $k^+$ and $k^-$ are the association and dissociation constants,
respectively, and $c$ the target concentration. The first term on the
right-hand side of Eq.~\eqref{eq:langmuir_ode_simple} is the hybridization
rate, which is partially controlled by the fraction $1-\theta$ of
available probes, whereas the second term is the denaturation rate. The
solution of Eq.~\eqref{eq:langmuir_ode_simple} with initial condition
$\theta(0)=0$ is
\begin{equation}\label{eq:langmuir_sol}
\theta(t) = \widetilde\theta \ (1 - e^{-t/\tau}),
\end{equation}
where $\tau \equiv (k^+c + k^-)^{-1}$ is the relaxation time and 
\begin{equation}\label{eq:langmuir_eq}
\widetilde\theta = \frac{cK}{1+cK}
\end{equation}
the value of $\theta$ at equilibrium, where we also introduced the
equilibrium constant, $K \equiv k^+/k^-$, of the reaction. The Langmuir
isotherm \eqref{eq:langmuir_eq} has been successfully employed in the past
for the description and quantification of DNA hybridization on a surface
at chemical equilibrium \cite{naef02, vain02, held03, zhan03, halp04,
bind05, burd06, carl06, nais09, hooy09, hadi12, harr13}. This relation
becomes linear in the target concentration, $\widetilde\theta \approx
cK$, when the probes are far from chemical saturation, \ie $cK \ll 1$
[or $\theta \ll 1$ in Eq.~\eqref{eq:langmuir_ode_simple}].

\subsubsection{Target depletion}

In the case of target depletion the hybridization kinetics is described by
\begin{equation}\label{eq:depletion_ode}
 \dod{\theta}{t} = k^+ (1 - \theta) ( c - a \theta ) - k^-\theta
=  a k^+ (\theta - \theta_+)(\theta - \theta_-),
\end{equation}
where $a$ is the probe concentration, and $\theta_\pm$ the 
two fixed points, given by
\begin{equation}\label{eq:depletion_1x1_roots}
 \theta_\pm = \frac{1}{2aK}\left[
 1+aK+cK \pm \sqrt{\left(1+aK+cK\right)^2-4acK^2}
 \right].
\end{equation}
Note that, the hybridization rate is now additionally controlled by
the amount of the remaining target in solution, \ie $c-a\theta$. Since
$\theta \le 1$, it follows that target depletion may be safely neglected
as long as $c \gg a$, \ie the initial target concentration is greater
than the probe concentration. Equation~\eqref{eq:depletion_ode} can
be solved through separation of variables. Using the initial condition
$\theta(0)=0$, one obtains
\begin{equation}\label{eq:depletion_1x1_solution}
\theta = \frac{\theta_+\theta_-(1-e^{-t/\tau})}{\theta_+-\theta_-e^{-t/\tau}},
\end{equation}
where the characteristic time now is $\tau \equiv [ ak^+ (\theta_+-\theta_-) 
]^{-1}$.  At long times $t \gg \tau$, the 
solution~\eqref{eq:depletion_1x1_solution} converges to $\theta_-$, which is a 
stable fixed point of Eq.~\eqref{eq:depletion_ode}, whereas $\theta_+$ is 
unstable. The approach to the stable fixed point is monotonic in $t$, as 
expected for a single first-order ordinary differential equation (ODE).  
Moreover, in the limit $a \to 0$, one finds $\theta_- = \widetilde{\theta}$ 
[given by Eq.~\eqref{eq:langmuir_eq}] and $\theta_+ \to \infty$. \edit{Finally, 
note that this equilibrium solution [smallest root in 
Eq.~\eqref{eq:depletion_1x1_roots}] is identical to Eq.~(6) of 
Ref.~\citenum{mich07} and Eq.~(7) of Ref.~\citenum{ono08}, apart from some 
constant factors.}

Equation~\eqref{eq:depletion_ode} may be generalized, so as to describe
the hybridization of $n_\text t$ different targets with $n_\text p$
different probes. The fraction $\theta_{ij}$ of the $i$-th probe
hybridized with the $j$-th target satisfies the differential equation
\begin{equation}\label{eq:depletion_ode_gen}
 \dod{\theta_{ij}}{t} = k^+_{ij} 
                        \bigg(1 - \sum_{m=1}^{n_\text t}\theta_{im}\bigg)
                        \bigg(c_j - \sum_{n=1}^{n_\text p}a_n\theta_{nj}\bigg)
                        -k^-_{ij}\theta_{ij}.
\end{equation}
Here $k^+_{ij}$ and $k^-_{ij}$ are the association and dissociation
constants, respectively, whereas $c_j$ and $a_n$ are the total
concentrations of the $j$-th target and the $n$-th probe, respectively.
Equations~\eqref{eq:depletion_ode_gen} constitute a set of coupled
nonlinear equations, which, in general, cannot be solved analytically. In
order to proceed, we will assume that the probes remain far from chemical
saturation \ie $\sum_{m=1}^{n_\text t}\theta_{im} \ll 1$, which leads
to the following set of linear equations
\begin{equation}\label{eq:depletion_ode_gen_lin}
\dod{\theta_{ij}}{t} \approx k^+_{ij} 
\bigg(c_j - \sum_{n=1}^{n_\text p}a_n\theta_{nj}\bigg)
-k^-_{ij}\theta_{ij}.
\end{equation}
The equations for $\theta_{ij}$ no longer couple the different targets
in solution (second index~$j$ in $\theta_{ij}$). As the spots are
not saturated, each target sequence has always probe sequences at its
disposal for hybridization, hence $\theta_{ij}$ and $\theta_{ij'}$
evolve independently from each other for $j \neq j'$. The equilibrium
hybridization fraction is given by (details are given in Appendix)
\begin{equation}\label{eq:depletion_eq}
\widetilde{\theta}_{ij} = 
\frac{c_j K_{ij}}{1+\sum_{n=1}^{n_\text p} a_n K_{nj}},
\end{equation}
where we have defined $K_{ij} \equiv k^+_{ij}/k^-_{ij}$, in analogy with
the case of a single probe/target pair.  For the numerical solution of
Eq.~\eqref{eq:depletion_ode_gen}, we used the Python implementation of
the LSODA ODE solver, using $10^4$ time steps. The latter were chosen
to be evenly spaced on a logarithmic scale (further supported by the
exponential nature of the solution), allowing for the accurate sampling of
both the short- and long-time behavior, while keeping the number of time
steps at a minimum. The kinetics can be solved analytically in the case
target depletion occurs due to a single probe, which is an interesting
case for application purposes.


\section{Results}

\begin{figure}[t]
\centering\includegraphics{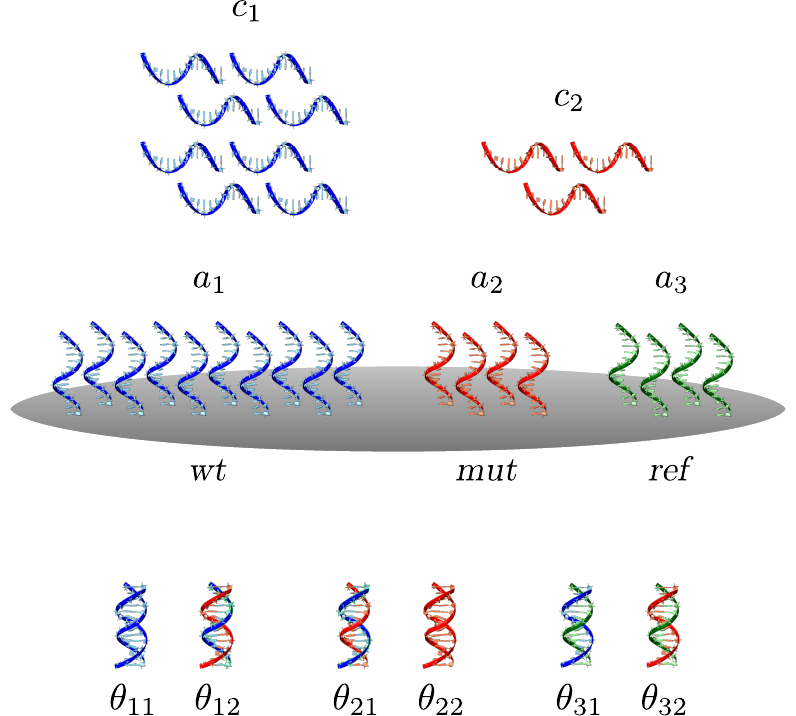}
\caption{A minimal experimental setup for the study of target
depletion. The sample solution (top) contains two targets, a wild-type
(blue) and a mutant sequence (red), with concentrations $c_1$ and $c_2$,
respectively. We assume the former to be in abundance, and the latter
to be present in small traces, \ie $c_1 \gg c_2$. The two targets come
into contact with three probes spotted on a surface (middle), with
concentrations $a_1$, $a_2$ and $a_3$. The first two probes (blue and
red) are the perfect complements of the two targets, while the third
probe (green) is used as a reference for the detection of the mutant
target. In order to achieve target depletion, we propose the use of a
large concentration, $a_1$, of wild-type probes. Finally, at the bottom
all possible duplexes are shown, together with the notation we employ.}
\label{fig:setup} 
\end{figure}

Here, we discuss the consequences of target depletion in conventional
hybridization experiments. In particular, we show that depletion can
significantly improve the performance of hybridization biosensors.
For this purpose, we consider the setup shown in Fig.~\ref{fig:setup},
which is simple enough to capture the basics of the process, yet, at
the same time, relevant for diagnostic applications: \edit{detecting small 
amounts of mutant DNA in a sample with a majority of wild type DNA.}

The sample in solution contains two types of target DNA, a wild-type
and a mutant sequence, the latter having a point mutation with respect
to the former. The two sequences have initial concentrations $c_1$ and
$c_2$, respectively. Particularly interesting for diagnostic purposes is
the detection of mutants at very low abundance, \eg $c_2/c_1 \ll 1$. To
this end, we employ a collection of three types of probes, a wild-type
({\sl wt}), a mutant ({\sl mut}) and a reference ({\sl ref}) probe,
immobilized on a surface at concentrations $a_1$, $a_2$ and $a_3$,
respectively (see Fig.~\ref{fig:setup}). While {\sl wt} and {\sl mut}
are the perfect complements of their target counterparts, {\sl ref}
contains contains one and two mismatches relative to the wild-type and mutant
targets, respectively. \edit{Further, the hybridization affinity of {\sl ref} to the 
wild-type target is designed to be equal to that of {\sl mut}. When a sample 
contains only wild-type target the equilibrated hybridization signal $\theta_3$ 
from {\sl ref} equals the $\theta_2$ from {\sl mut}, hence the name reference probe.} 
\edit{When the sample also contains a trace of mutant target DNA, $\theta_2$ but not 
$\theta_3$ will be significantly affected.}

\edit{More quantitively,} the signal measured from each probe is the sum of the contributions from the wild type and the mutant, 
\ie $\theta_i = \theta_{i1} + \theta_{i2}$. 
Probes {\sl mut} and {\sl ref} both have a single
mismatch with respect to the wild-type target. 
\edit{By design we assume that their hybridization affinity to the wild type is similar, 
hence} $\theta_{21} \approx \theta_{31}$, which can be achieved with
a proper choice of the reference probe. In case a mutant target is
present in solution ($c_2>0$), one has $\theta_{22} \gg \theta_{32}$,
due to its much higher affinity for the second probe (perfect complement)
than the third probe (two mismatches). Following Ref.~\citenum{will17},
we define the detection signal as
\begin{equation}\label{eq:det_signal}
S \equiv \log\frac{\theta_2}{\theta_3}
= \log\frac{\theta_{21} + \theta_{22}}{\theta_{31} + \theta_{32}}
\approx \log\left(1 + \frac{\theta_{22}}{\theta_{21}}\right),
\end{equation}
which will be zero when $c_2 = 0$ ($\theta_{22} = 0$) and positive
otherwise.  Clearly, for diagnostic purposes we wish to have a large value
of $S$ for small ratios $c_2/c_1$.  Note that cross-hybridization can
cause $\theta_{21}$ to be much larger than $\theta_{22}$, especially
when $c_1 \gg c_2$, hence obscuring the detection of the mutant
target. In order to address this issue, we propose the use of a large
concentration~$a_1$ of {\sl wt}, which will deplete the corresponding
target, hence leading to a cleaner signal from {\sl mut}. Though
perhaps evident, this approach will also deplete the mutant target, and
a profound quantitative analysis is needed to investigate this issue.
In what follows, we will quantify this effect by considering both the
equilibrium and kinetics of the hybridization process.

\subsection{Equilibrium properties}%

We will first focus on the equilibrium aspects of target depletion. For
a system with three probes, the hybridized probe fraction at equilibrium
is given by [see Eq.~\eqref{eq:depletion_eq}]
\begin{equation}
\widetilde{\theta}_{2j} = \frac{c_j K_{2j}} 
{1 + a_1 K_{1j} + a_2 K_{2j} + a_3 K_{3j}}.
\end{equation}
The detection signal is then given by
\begin{equation}
S = \log\left( 1 + \frac{\widetilde{\theta}_{22}}{\widetilde{\theta}_{21}} 
\right) 
\approx  \log\left( 1 + \frac{c_2}{c_1} \, \frac{K_\PM}{K_\MM} \,
\frac{1 + a_1 K_\PM + a_2 K_\MM + a_3 K_\MM} {1 + a_1 K_\MM + a_2 K_\PM + 
a_3 K_\MMM} \right).
\end{equation}
For simplicity, we have assumed that $K_{11} = K_{22} \equiv K_\PM$,
$K_{12} = K_{21} = K_{31} \equiv K_\MM$ and $K_{32} \equiv K_\MMM$,
associated with the perfect-match, single-mismatch and two-mismatch
hybridizations, respectively.  It is important to stress that the
above relations are introduced for convenience and do not affect the
main conclusions of this work. In absence of depletion ($a_i=0$), the
detection signal becomes
\begin{equation}\label{eq:signal_langmuir_eq}
 S_0 = \log \left(1 + \frac{c_2}{c_1} \frac{K_\PM}{K_\MM} \right)
     = \log \left(1 + \frac{c_2}{c_1} e^{\Delta\Delta G_\MM/RT} \right),
\end{equation}
where we have used the thermodynamic relation $K = e^{-\Delta G/RT}$,
with $\Delta G$ the hybridization free energy, $R$ the gas constant
and $T$ the temperature (note that by convention $\Delta G<0$). We have
also introduced $\Delta\Delta G_\MM \equiv \Delta G_\MM - \Delta G_\PM$,
the free-energy difference between the perfect-match and one-mismatch
hybridizations which depends on the mismatch identity and on flanking
nucleotides, according to the nearest-neighbor model of DNA hybridization
\cite{sant04, hooy09, hadi12}.  Equation~\eqref{eq:signal_langmuir_eq}
has been experimentally verified in the past (see \eg Fig.~3 of
Ref.~\citenum{will17}), and shows that there are two factors controlling
the detection limit of the device. One is the relative abundance,
\ie it is easier to detect mutants at high relative concentrations
($c_2/c_1$). The other factor is the relative affinity $\Delta \Delta
G_\MM > 0$, \ie a large free energy penalty for mismatched hybridization
leads to suppression of cross hybridization, and hence facilitates the
detection of the mutant. Since typical values of $\Delta\Delta G_\MM$
lie in the range $1-4$~kcal/mol~\cite{hooy09}, and using the fact that the
signal is detectable only when $S \ge S_\text{min} = 0.5$~\cite{will17},
it follows that the minimum relative concentration, \edit{$c_2/c_1$}, that can 
be measured with this method lies in the range \edit{0.17\% to 15\%,}
in agreement with previous reports~\cite{will17}. In this calculation we used 
$T = 65^\circ$C as a typical system temperature~\cite{hooy09, hooy10a, will17}.

Next, we consider the other limit of strong depletion. We will assume
the target depletion to occur only due to the wild-type probe, which
can be tuned by choosing a large-enough probe concentration so
that the condition $a_1 K_\MM \gg a_2 K_\PM$ is met. Moreover, by fully
exploiting the effect of target depletion, so that $a_1 K_\MM \gg 1$,
we obtain the following elegant expression
\begin{equation}\label{eq:signal_depletion_eq}
 S \approx \log\left[1+\frac{c_2}{c_1} \left(\frac{K_\PM}{K_\MM}\right)^2 
\right]
   = \log\left(1+\frac{c_2}{c_1} e^{2\Delta\Delta G_\MM/RT}\right).
\end{equation}
By comparing Eqs.~\eqref{eq:signal_langmuir_eq} and
\eqref{eq:signal_depletion_eq}, we see that depleting the abundant
wild-type target indeed leads to higher $S$ (additional factor of two in
the exponent).  Performing the same analysis as above, we find that the
minimum relative concentration, \edit{$c_2/c_1$}, that is experimentally 
detectable is in the range \edit{0.00044\% to 3.3\%.}
This corresponds to an 
enhancement
of the detection sensitivity by one to three orders of magnitude, owing
to target depletion.

In order to experimentally realize the aforementioned detection
enhancement, two conditions need to be met, as mentioned above. First,
the relative concentration $a_1$ of the wild-type probes has to be much
larger than those of the mutant and reference probes, so that
\begin{equation}\label{eq:condition1}
\frac{a_1}{a_{n}} \gg \frac{K_\PM}{K_\MM} = e^{\Delta\Delta G_\MM/RT},
\end{equation}
with $n=2,3$. Using typical values for hybridization free energies
of single base pair mismatches (see above) we estimate $4 \lesssim
\exp(\Delta\Delta G_\MM/RT) \lesssim 400$.  Thus, the larger the
free-energy penalty, $\Delta\Delta G_\MM$, of a mismatch is, the larger
the ratio $a_1/a_n$ ($n=2,3$) needed. Moreover, the absolute value of
$a_1$ needs to be large enough, so as to maximize the contribution of
depletion. The precise condition is
\begin{equation}\label{eq:condition2}
 a_1 \gg \frac{1}{K_\MM} = e^{\Delta G_\MM/RT}.
\end{equation}
The precise value of $\Delta G_\MM/RT$ depends strongly on the DNA
sequence, and can be estimated based on the nearest-neighbor model of
DNA~\cite{sant04}.

\edit{As a practical evaluation of these predictions, we also performed a DNA 
microarray experiment. The microarray contained two drastically-different 
groups of sequences, allowing us to study DNA hybridization both in absence and 
presence of target depletion (see Appendix for details). The solution contained 
both a wild-type and a mutant target, with relative proportion equal to 
$c_2/c_1 = 5\%$. In absence of depletion, the observed detection signal, as 
defined in Eq.~\eqref{eq:det_signal}, was found to be $S_0 = 0.9 \pm 0.1$. 
On the other hand, target depletion was found to yield the value $S = 4.5 \pm 
0.2$, corresponding to a five-fold enhancement of the detection signal 
(See details in Appendix). 
Interestingly, by taking $\Delta\Delta G_\MM = 2.4$~kcal/mol, which is a 
reasonable value for the cost of a mismatch~\cite{hooy09}, 
Eqs.~\eqref{eq:signal_langmuir_eq} and \eqref{eq:signal_depletion_eq} yield 
$S_0 \approx 1.0$ and $S \approx 4.2$, respectively. These values are 
remarkably close to the experimental ones, given the presence of a single free 
parameter.}

\edit{Performing again the same analysis as above this leads to an estimated detection limit of the relative concentration $c_2/c_1$ of 1.8\% for the non-depletion case and 0.052\% for the depletion case. Hence the sensitivity is increased by a factor of 35 through target depletion.}

\subsection{Hybridization Kinetics}%

To investigate the kinetics of the system, we have numerically solved the
coupled ODE~\eqref{eq:depletion_ode_gen}. The wild-type concentration was
fixed at the experimentally-realistic value of $c_1 = 50$~pM, while to
obtain strong depletion we have set $a_1 = 800$~pM and $a_2 = a_3 = 4$~pM.
We considered on-rates identical for all sequences, which is supported
to a good extent by experimental observations~\cite{glaz06}. The value
was set to $k_+=10^6$~$\text{s}^{-1} \text{M}^{-1}$. The off-rates were
then fixed by the equilibrium condition $K \equiv k_+/k_- = e^{-\Delta
G/RT}$. For the perfect-match, one- and two-mismatch hybridizations we
used $\Delta G_\PM = -16$~kcal/mol, $\Delta G_\MM = -13.5$~kcal/mol and
\edit{$\Delta G_\MMM = -11$~kcal/mol},
respectively, based on nearest-neighbors
data for a 15-mer at $T = 65^\circ$C \cite{sant04}.

\begin{figure}[t]
\centering\includegraphics{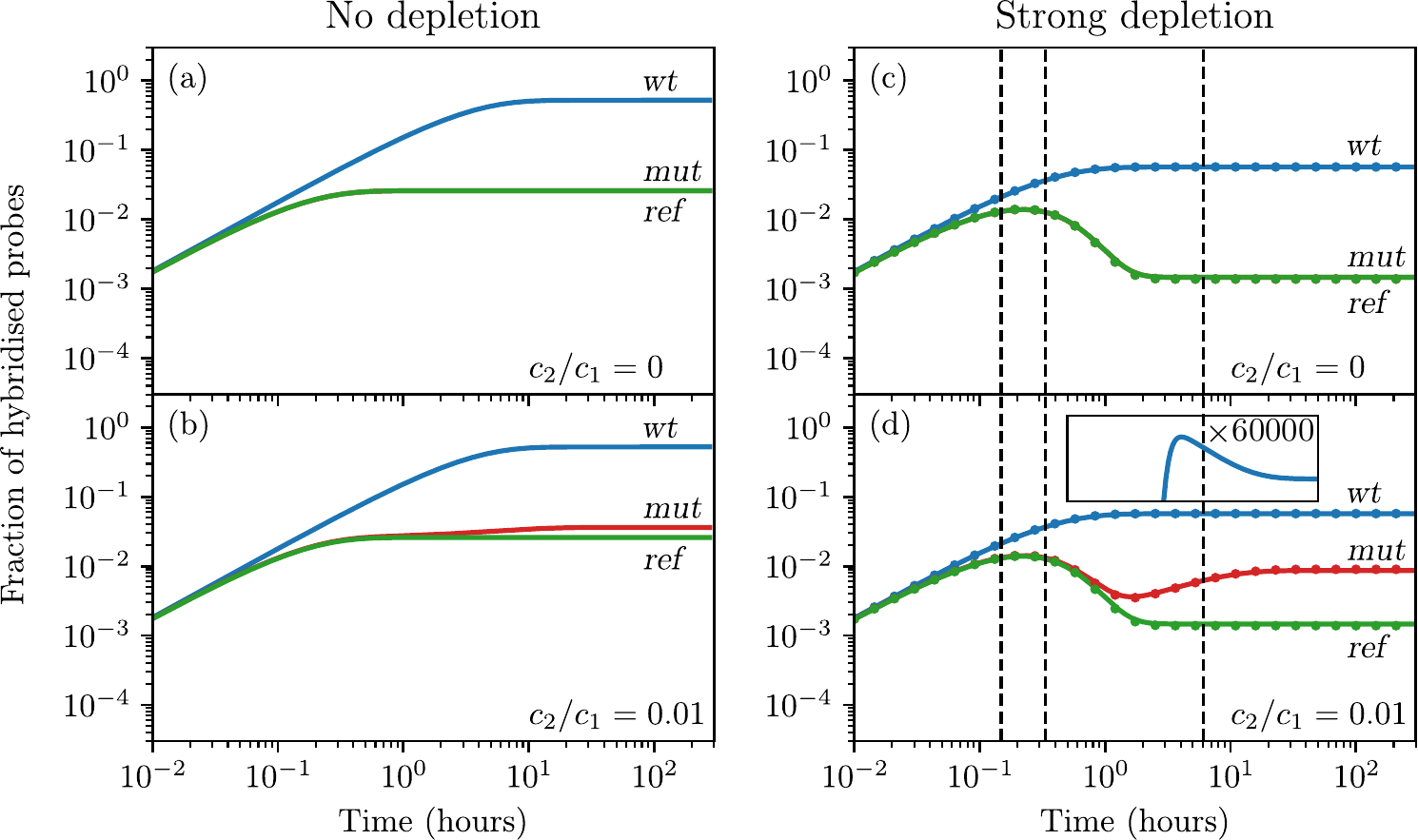}
\caption{Hybridization evolution of a wild-type and a mutant target with
an array of three probes, a wild-type (blue line), a mutant (red line)
and a reference one (green line), for two values of the relative target
concentration $r \equiv c_2/c_1$. Panels (a) and (b) correspond to the
case where no depletion of the wild-type target takes place, whereas (c)
and (d) to the strong depletion case, where the wild-type probe is in
excess concentration. In the latter case, besides the numerical solution
of Eqs.~\eqref{eq:depletion_ode_gen} (solid lines), we also plot the
analytical solution~\eqref{eq:twotargets_rates} (points). The dashed
vertical lines correspond to the three characteristic times, $t_1$,
$t_2$ and $t_3$, discussed in the main text. The inset in (d) shows a
magnification of the {\sl wt} signal, revealing a very small overshoot.
\label{fig:3probes_evolution}}
\end{figure}

Figure~\ref{fig:3probes_evolution} shows the hybridization kinetics
for $\theta_i$ in the case of no-depletion (a and b) and of strong
depletion (c and d). The \edit{dots}
are the analytical solution, while the \edit{solid lines}
are obtained from 
the numerical integration of
Eq.~\eqref{eq:depletion_ode_gen}. In (a) and (c) the solution contains
wild-type target at concentration $c_1=50$~pM and no mutant ($c_2=0$). The
signals measured from {\sl mut} and {\sl ref} perfectly overlap,
since we have assumed equal hybridization free energy $\Delta G_\MM$
for the two sequences. In (b) and (d) the solution additionally contains
$c_2=0.5$~pM mutant (corresponding to a ratio $c_2/c_1 = 0.01$, \ie $1\%$
of the wild-type concentration). Figure~\ref{fig:3probes_evolution}
indicates that the presence of the mutant is more easily detectable in
the case of strong depletion as the gap between {\sl mut} and {\sl ref}
is much more pronounced. The kinetics is also remarkably different:
in absence of depletion the signal increases monotonically in time,
whereas in the strong depletion case we observe a nonmonotonic behavior
and even a dip in the signal from {\sl mut}.

The long time behavior shown in Fig.~\ref{fig:3probes_evolution}
corresponds to the equilibrium solution given by
Eq.~\eqref{eq:depletion_eq}. In order to understand the observed rich
kinetics, one can use a simplified solvable case in which we assume that
the depletion occurs due to the {\sl wt} probe alone, \ie $a_2,a_3
\approx 0$ and $a_1 \equiv a \neq 0$.  Under this approximation,
the solution $\theta_i = \theta_{i1}+\theta_{i2}$ is found to be [see
Eq.~\eqref{eq:sol_probei} of Appendix]
\begin{equation}\label{eq:twotargets_rates}
 \begin{aligned}
  \theta_1 &= \frac{c_1K_\PM}{1+aK_\PM} \left[ 1 - e^{-(ak_++k_\PM)t} \right]
              +\frac{c_2K_\MM}{1+aK_\MM} \left[ 1 - e^{-(ak_++k_\MM)t} \right], 
              \\[10pt]
  \theta_2 &= \frac{c_1K_\MM}{1+aK_\PM} \left\{ 1 - e^{-k_\MM t} + 
\frac{K_\PM}{K_\MM} \frac{ak_+}{ak_+ + k_\PM - k_\MM} \left[ e^{-k_\MM t} - 
e^{-(ak_++k_\PM)t} \right] \right\} \\
           &+ \frac{c_2K_\PM}{1+aK_\MM} \left\{ 1 - e^{-k_\PM t} + 
\frac{K_\MM}{K_\PM} \frac{ak_+}{ak_+ + k_\MM - k_\PM} \left[ e^{-k_\PM t} - 
e^{-(ak_++k_\MM)t} \right] \right\},\\[10pt]
  \theta_3 &= \frac{c_1K_\MM}{1+aK_\PM} \left\{ 1 - e^{-k_\MM t} + 
\frac{K_\PM}{K_\MM} \frac{ak_+}{ak_+ + k_\PM - k_\MM} \left[ e^{-k_\MM t} - 
e^{-(ak_++k_\PM)t} \right] \right\} \\
           &+ \frac{c_2K_\MMM}{1+aK_\MM} \left\{ 1 - e^{-k_\MMM t} + 
\frac{K_\MM}{K_\MMM} \frac{ak_+}{ak_+ + k_\MM - k_\MMM} \left[ e^{-k_\MMM t} - 
e^{-(ak_++k_\MM)t} \right] \right\},
 \end{aligned}
\end{equation}
where we used $k_\PM$, $k_\MM$ and $k_\MMM$ to denote the off-rates,
while $K_\PM = k_+/k_\PM$, $K_\MM = k_+/k_\MM$ and $K_\MMM =
k_+/k_\MMM$.  Equations~\eqref{eq:twotargets_rates} are shown in
Fig.~\ref{fig:3probes_evolution} as dotted lines and are in excellent
agreement with numerics. One can further simplify them using the
assumption $aK_\PM > aK_\MM \gg 1$, which corresponds to the limit
of strong depletion. This condition is satisfied for the parameters
used in Fig.~\ref{fig:3probes_evolution}. Under this assumption,
Eqs.~\eqref{eq:twotargets_rates} reduce to
\begin{equation}\label{eq:twotargets_rates_app}
\begin{aligned}
  \theta_1 &\approx \frac{c_1+c_2}{a} \left( 1 - e^{-ak_+t} \right), \\[10pt]
  \theta_2 &\approx \frac{c_1K_\MM}{aK_\PM} \left[ 1 - e^{-k_\MM t} + 
\frac{K_\PM}{K_\MM} \left( e^{-k_\MM t} - e^{-ak_+t} \right) \right] \\
           &+ \frac{c_2K_\PM}{aK_\MM} \left[ 1 - e^{-k_\PM t} + 
\frac{K_\MM}{K_\PM} \left( e^{-k_\PM t} - e^{-ak_+t} \right) \right],\\[10pt]
  \theta_3 &\approx \frac{c_1K_\MM}{aK_\PM} \left[ 1 - e^{-k_\MM t} + 
\frac{K_\PM}{K_\MM} \left( e^{-k_\MM t} - e^{-ak_+t} \right) \right].
 \end{aligned}
\end{equation}
In the last expression we have neglected the contribution $\theta_{32}$
of the mutant target to the probe {\sl ref}, as the corresponding
hybridization involves two mismatches and $c_2 \ll c_1$. We, thus,
identify three characteristic times, $t_1 \equiv 1/ak_+$, $t_2 \equiv
1/k_\MM$ and $t_3 \equiv 1/k_\PM$, which are ordered as $t_1 < t_2 <
t_3$ and are shown as dashed vertical lines in panels (c) and (d)
of Fig.~\ref{fig:setup}. We note that, while $\theta_1$ is clearly a
monotonic function of time, there are several time-dependent factors with
opposite signs in $\theta_2$ and $\theta_3$, giving rise to nonmonotonic
time evolution.

To analyze this time dependence in more detail, we first consider the
regime in which $t \lesssim t_1$. In this time interval we approximate
$\exp(-a k_+ t) \approx 1 - a k_+ t$ and $\exp(-k_\MM t) \approx
\exp(-k_\PM t) \approx 1$, so as to get
\begin{equation}
\theta_1 \approx \theta_2 \approx \theta_3 \approx 
\left( c_1+c_2 \right) k_+ t,
\end{equation}
which indicates that at short time scales the kinetics is characterized
by an identical binding rate to {\sl wt}, {\sl mut} and {\sl ref}. This
is because we have assumed equal attachment rate $k_+$ for all probes
and targets, which is a reasonable approximation.  This, however, does
not influence the main features of the kinetics at the subsequent time
scales. In the next interval $t_1 \ll t \lesssim t_2$, we approximate
$\exp(-a k_+ t) \approx 0$ and $\exp(-k_\PM t) \approx 1$. In this case
the {\sl wt} probe signal reaches a stationary value $\theta_1 \approx
(c_1+c_2)/a$, which can also be obtained from the equilibrium solution
\eqref{eq:depletion_eq}, while
\begin{equation}
\begin{aligned}
\theta_2 &\approx \frac{c_1}{a} \left[ \frac{K_\MM}{K_\PM} + \left( 1 - 
                   \frac{K_\MM}{K_\PM} \right) e^{-k_\MM t} \right]
                   + \frac{c_2}{a}, \\[10pt]
\theta_3 &\approx \frac{c_1}{a} \left[ \frac{K_\MM}{K_\PM} + \left( 1 - 
                   \frac{K_\MM}{K_\PM} \right) e^{-k_\MM t} \right],
\end{aligned}
\end{equation}
which, as $K_\PM > K_\MM$, are decreasing functions of time. In this
regime the wild-type target starts dissociating at the same rate
$k_\MM$ from {\sl mut} and {\sl ref} probes. This leads to a very
weak increase in the hybridization of the {\sl wt} probe, which is
not detectable in the scale of Fig.~\ref{fig:3probes_evolution} (see
inset of panel d), and also not present in the approximated solution
\eqref{eq:twotargets_rates_app}. This weak increase is, however,
present in the full solution \eqref{eq:twotargets_rates}.  Finally, at
even longer times, \ie for $t_2 \ll t \sim t_3$, one has $\exp(-a k_+
t) \approx  \exp(-k_\MM t) \approx 0$. The {\sl ref} probe reaches a
steady state $\theta_3 \approx c_1 K_\MM/ ( aK_\PM)$, while the {\sl mut}
probe increases monotonically as
\begin{equation}
\begin{aligned}
\theta_2 &\approx \frac{c_1}{a} \frac{K_\MM}{K_\PM}
                  + \frac{c_2}{a} \left[ \frac{K_\PM}{K_\MM} - \left( 
                  \frac{K_\PM}{K_\MM} - 1 \right) e^{-k_\PM t} \right].
\end{aligned}
\end{equation}
This increase takes place only if $c_2 >0$, while in absence of mutant
target ($c_2=0$) this third timescale is absent, and {\sl mut} reaches
a steady state value from above as {\sl ref}. In this last regime
the wild-type target has completely equilibrated, and the mutant
target gets redistributed from the {\sl wt} probe to the {\sl mut}
probe. This gives rise to a monotonic increase in the hybridization
of the latter, until the complete equilibration of the system. The
turnover time at which $\theta_2$ is minimal can be calculated from
Eqs.~\eqref{eq:twotargets_rates_app} and is given by
\begin{equation}\label{eq:tmin}
 t_\text{min} = \frac{\log(c_1/c_2)}{k_\MM-k_\PM}.
\end{equation}

\begin{figure}[t]
\centering\includegraphics{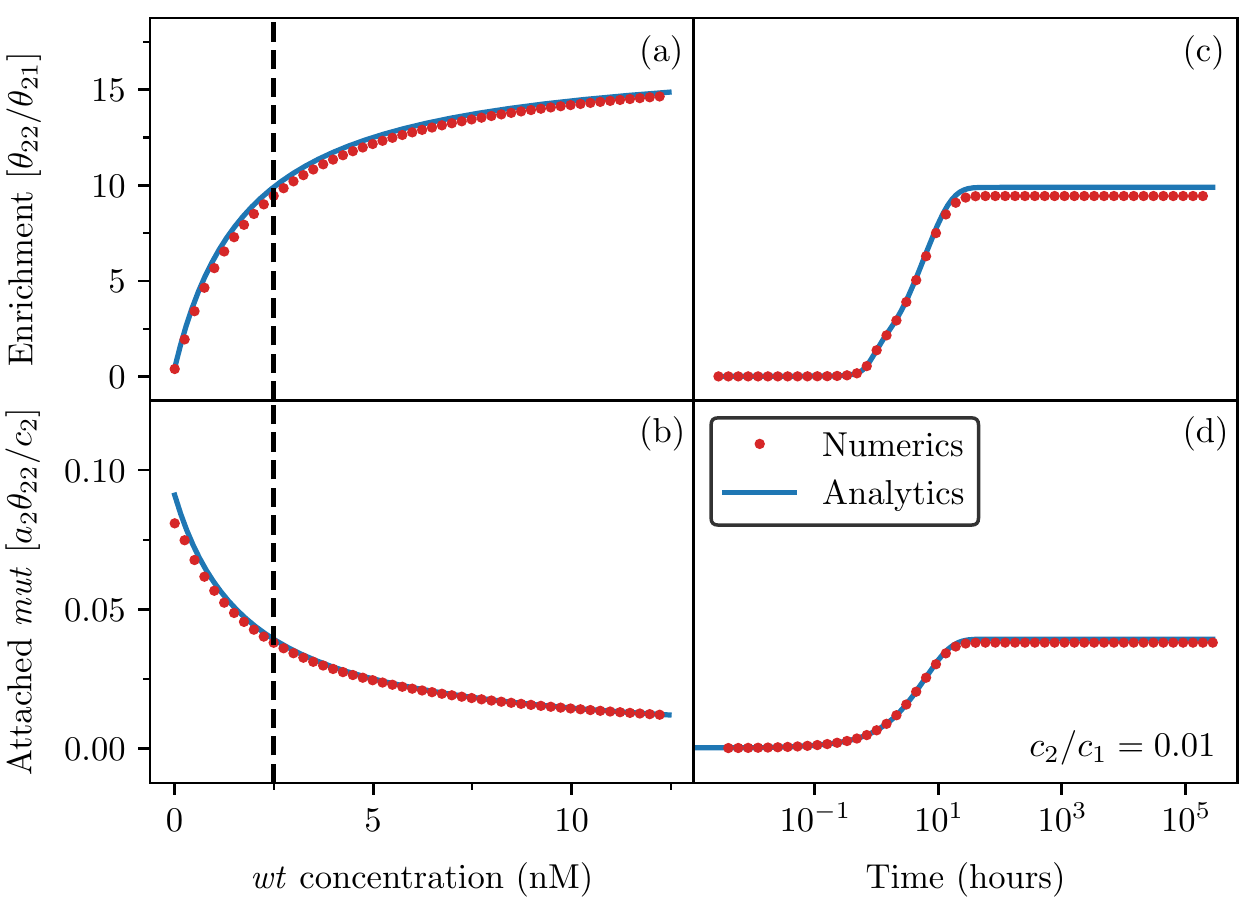}
\caption{Upper panels: Mutant/wild-type target ratio,
$\theta_{22}/\theta_{21}$, attached to the \textit{mut} probe. Higher
values of the ratio enable the further enrichment of the sample, by
increasing the relative population of the mutant with respect to the
wild-type target. Lower panels: Fraction, $a_2\theta_{22}/c_2$, of the
mutant target that has hybridized with the \textit{mut} probe. For
application purposes, the concentration, $a_2\theta_{22}$, of the
captured mutant target should be comparable to the initial one, $c_2$,
in solution. These quantities are plotted both (a,b) as a function
of the \textit{wt} probe concentration at equilibrium and (c-d) as a
function of time at a fixed \textit{wt} \edit{probe} concentration ($a_1 = 
2.5$~nM,
corresponding to the dashed vertical lines).  The small deviation between
analytics and numerics arises from the approximation $a_2, a_3 \approx 0$
included in the former.
\label{fig:enrichment}}
\end{figure}

Next, we show how target depletion can be used for sample enrichment,
\ie increasing the relative amount of mutant DNA over wild-type
DNA. From an application point of view, this is an important issue,
and can lead to an increased performance for mutant detection by other
techniques, such as sequencing~\cite{samo15, khod16}. Hereto, we focus
on the hybridized material on the \textit{mut} probe (probe number 2)
and study two important quantities (see Fig.~\ref{fig:enrichment}): the
ratio of mutant over wild-type target attached to \textit{mut} (a,c)
and the absolute amount of mutant target (b,d).  The former quantity
determines whether we can achieve enrichment, the latter is needed as a
measure of capture efficiency. In Fig.~\ref{fig:enrichment} (a) and (b)
these quantities are plotted as functions of depletion (\ie \textit{wt} 
\edit{probe} concentration, $a_1$), for a sample with starting target mutant 
ratio of
$c_2/c_1 = 0.01$.  These plots quantitatively show how depletion leads to
a trade-off between yield and sample enrichment. As an example, indicated
with the dashed vertical line, is a regime where a yield of about 4\%
gives a mutant enrichment of a factor $(\theta_{22}/\theta_{21})/(c_2/c_1)
\approx 940$.  Finally, the kinetics shown in Fig.~\ref{fig:enrichment}
(c) and (d) indicates that the quantities evolve monotonically in time,
hence equilibrium conditions can be used to achieve the best results.

\section{Conclusion}

In this paper we have analyzed the equilibrium and kinetics of hybridization in 
DNA biosensors under the effect of strong target depletion.  This is a condition 
which has been considered only in \edit{limited prior studies~\cite{mich07, 
ono08, burd09}} since the typical assumption behind hybridization models in DNA 
biosensors, as the Langmuir adsorption model, is that the target sequences in 
solution are in excess. Target concentration is then considered to be constant 
throughout the duration of the experiment. To fulfill this condition one needs a 
sufficient amount of hybridizing material to start with.  Although target 
depletion is typically avoided, our analysis shows that one can turn it, in some 
applications, into an advantageous condition, leading to an increase of the 
performance of the biosensor.

We focused on the problem of detection of small amount of mutant sequence
(with concentration $c_2$) diluted in a highly-abundant wild type
(with concentration $c_1$), and specifically addressed the case of a
single nucleotide difference between the two. An example where this is
an important diagnostic problem is in liquid biopsy, where one examines a
mixture of ``healthy'' molecules in majority, with a small subpopulation
of molecules carrying a specific pathogenic property.

The minimal design employed in this study involves three probe sequences,
which we referred to as wild-type ({\sl wt}), mutant ({\sl mut}) and
reference ({\sl ref}) probe. The presence of the mutant in solution
is inferred by the ratio of hybridization signals measured from {\sl
ref} and {\sl mut}. We have presented a quantitative analysis of
the hybridization kinetics and shown that in presence of wild-type
depletion one can decrease the detection limit up to three orders of
magnitudes in the ratio $c_2/c_1$ [as revealed by a comparison between
Eqs.~\eqref{eq:signal_langmuir_eq} and \eqref{eq:signal_depletion_eq}].
Note that the only sequence-dependent parameter controlling the detection
limit is the free energy penalty associated to a single mismatch. With
the same design we showed that, next to detection, also target enrichment
can be enhanced.

Finally, our analysis of the kinetics revealed a rich behavior, with
interplay between the initial strong binding of target, followed by
unbinding and redistribution of the sequences between the different
probes. This resulted in three different time scales and a {\sl mut}
signal that exhibits a nonmonotonic behavior: an increase followed by
a decrease and then by a final increase towards equilibrium.  We expect
that this distinct feature, which we have found to take place only when
$c_2 > 0$, should be observable in experiments which have access to the
kinetics of hybridization.~\cite{verm07, van11}


\begin{acknowledgement}

We acknowledge financial support from the Research Funds Flanders (FWO
Vlaanderen) Grant No.\ VITO-FWO 11.59.71.7N.

\end{acknowledgement}




\appendix
\section{Appendix} 
\subsection{\edit{DNA microarray experiment}}

\edit{To confirm the detection enhancement predicted by 
Eqs.~\eqref{eq:signal_langmuir_eq} and \eqref{eq:signal_depletion_eq}, we 
performed a microarray experiment. We designed two sets of wild-type, mutant and 
reference probes, shown in Table \ref{table:probes}. The first probe set was 
based on previously-published data~\cite{will17}, from which we selected the 
optimal double-mismatch {\sl ref} probe for the {\sl wt} and {\sl mut} pair, \ie 
one for which $|\theta_{21} - \theta_{31}|$ is minimised. (Note that the selected {\sl ref} probe 
which best fulfilled this criteron actually contains two mismatches with respect to the wild-type target 
and three to the mutant target, in contrast to the one- and two-mismatch case described in the main text. 
The number of mismatches is not critical here, but rather the relative $\Delta \Delta G$ values.)}

\edit{The second probe set was designed to maintain the sequences and positions of the variable triplets 
from the first set, as well as a similar overall $\Delta G_\PM$, while the 
remainder of the sequence was kept distinct to avoid cross-hybridisation.}

\begin{figure}[t]
\centering\fedit{\includegraphics{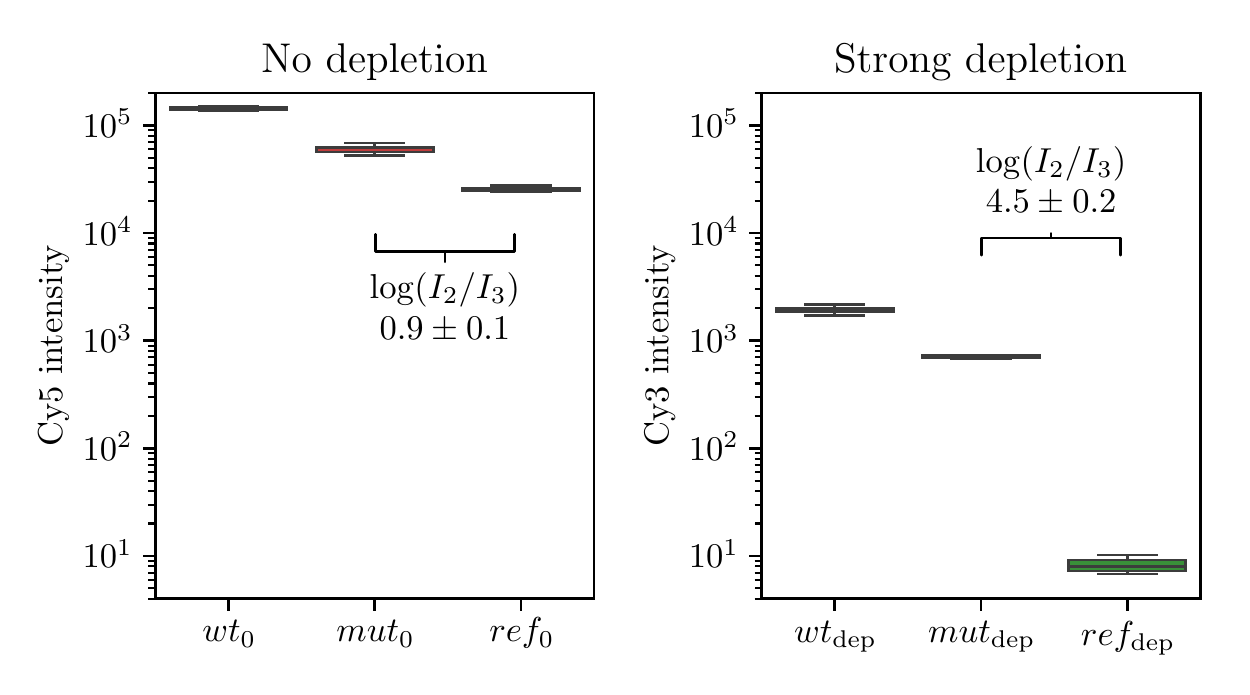}}
\caption{\edit{Microarray fluorescence intensity data showing the effect of depletion 
on the relative mutant hybridisation signal.}
\label{fig:4experiment}}
\end{figure}

\edit{The six probes were laid out on the array, which contained 10 spots of 
the $wt_\undep$ probe and $1.5\times 10^4$ spots of the $wt_\dep$ probe, 
corresponding to the no depletion and strong depletion regimes. The array was 
incubated with a mixture of $wt$ and $mut$ targets for each probe set (Table 
\ref{table:targets}), with $c_2/c_1 = 0.05$. Targets for the two probe sets 
were labelled with different fluorophores, allowing them to be measured 
independently on the same array. The results are shown in Figure \ref{fig:4experiment}}.

\begin{table}[]
\fedit{
\begin{tabular}{llll}
\hline
                              & Probe  & Sequence (5' - 3') & Array replicates \\ \hline
\multirow{3}{*}{Depletion}    & $wt_\dep$    & \texttt{GTTG\wtriplet{GAG}CT\wtriplet{GGT}GGC\wtriplet{GTA}GGCAA} & 15158                 \\
                              & $mut_\dep$   & \texttt{GTTG\wtriplet{GAG}CT\mtriplet{GCT}GGC\wtriplet{GTA}GGCAA} & 10                    \\
                              & $ref_\dep$   & \texttt{GTTG\mtriplet{GGG}CT\wtriplet{GGT}GGC\mtriplet{GAA}GGCAA} & 10                    \\ \cline{2-4} 
\multirow{3}{*}{No depletion} & $wt_\undep$  & \texttt{CGCC\wtriplet{GAG}TC\wtriplet{GGT}CAT\wtriplet{GTA}CTGGC} & 10                    \\
                              & $mut_\undep$ & \texttt{CGCC\wtriplet{GAG}TC\mtriplet{GCT}CAT\wtriplet{GTA}CTGGC} & 10                    \\
                              & $ref_\undep$ & \texttt{CGCC\mtriplet{GGG}TC\wtriplet{GGT}CAT\mtriplet{GAA}CTGGC} & 10                    \\ \cline{1-4} 
\end{tabular}} 
\caption{\edit{Microarray probe sequences used in the experiment.}}
\label{table:probes}
\end{table}

\begin{table}[]
\fedit{
\begin{tabular}{llll}
\hline
                              & Target  & Sequence  (5' - 3') & Concentration \\ \hline
\multirow{2}{*}{Depletion}    & $\targetwt{\dep}$    & \texttt{TTGCCTACGCC\wtriplet{ACC}AGCTCCAAC}+Cy3 & 95 pM       \\
                              & $\targetmut{\dep}$   & \texttt{TTGCCTACGCC\mtriplet{AGC}AGCTCCAAC}+Cy3 & 5 pM       \\ \cline{2-4} 
\multirow{2}{*}{No depletion} & $\targetwt{\undep}$  & \texttt{GCCAGTACATG\wtriplet{ACC}GACTCGGCG}+Cy5 & 95 pM       \\
                              & $\targetmut{\undep}$ & \texttt{GCCAGTACATG\mtriplet{AGC}GACTCGGCG}+Cy5 & 5 pM       \\ \cline{1-4} 
\end{tabular}}
\caption{\edit{Target sequences used in the microarray experiment.}}
\label{table:targets}
\end{table}

\subsubsection{\edit{Materials and methods}}

\edit{A custom 8x15K Agilent microarray slide was used (Agilent Technologies, 
US). Microarray probes all included a $(\text{dA})_{30}$ linker sequence on the 
3' end. Target oligonucleotides (IDT, Germany) included a $(\text{dA})_{20}$ 
linker and a Cy3 or Cy5 fluorescent dye on the 3' end. Target oligonucleotides 
were mixed to final concentrations shown in Table \ref{table:targets} in $1\times$ 
Agilent GEx hybridisation buffer HI-RPM with $1\times$ Agilent GE blocking agent. The 
slide was incubated with 40 \si{\micro\liter} of this target mixture in an 
Agilent hybridisation oven for 17 hours at 65\si{\degreeCelsius} with rotor 
setting 10, followed by washing according to manufacturer instructions. An 
Agilent G2565BA scanner was used to image the microarray, with a 
5-\si{\micro\meter} resolution and 100\% gain. Image analysis was carried out 
using the Agilent Feature Extraction software, version 10.7. The signal was 
background-corrected by subtraction of the global minimum signal.}

\subsection{Equilibrium isotherm and the Sherman-Morrison formula}

In order to compute the fraction of hybridized probes at equilibrium, it
is convenient to introduce a vector $\vecs\theta$, defined as $\vecs\theta
\equiv \{ \theta_{11}, \theta_{21}, \ldots, \theta_{n_\text p 1},
\theta_{12}, \theta_{22}, \ldots, \theta_{n_\text p 2}, \ldots \}$. With
this definition, one can cast Eq.~\eqref{eq:depletion_ode_gen_lin}
in matrix form
\begin{equation}\label{eq:ode_mat}
\dod{\vecs\theta}{t} = - \vec M \vecs\theta + \vec b,
\end{equation}
where $\vec M$ is a block diagonal matrix. The j-th block, $\vec M^j$,
corresponds to the contribution from a single target~$j$ and mixes
the elements of the subvector ${\vecs\theta}^j \equiv \{ \theta_{1j},
\theta_{2j}, \ldots, \theta_{n_\text p j}\}$. Its entries are
\begin{equation}
M^j_{nm} = k^+_{nj} a_m + k^-_{nj} \delta_{nm},
\label{defM}
\end{equation}
where $\delta_{nm}$ indicates the Kronecker $\delta$ function. In the
$j$-th block, the constant vector is given by $\vec{b}^j_n = k^+_{nj}
c_j$. The equilibrium value $\widetilde{\vecs{\theta}}$ is obtained by
inverting the matrix $\vec{M}$ as
\begin{equation}
\widetilde{\vecs{\theta}} = \vec{M}^{-1} \vec b,
\end{equation}
which can be performed independently for each block. In order to calculate
$\vec M^{-1}$, we notice that the j-th block of $\vec M$ is the sum of a
diagonal matrix and the outer product of two vectors, \ie $\vec{M}^j =
\vec{D} + \vec{u} \vec v^\text T$, with $\vec D$ diagonal and $\vec u
\vec v ^\text T \equiv \vec u \otimes \vec v$. This allows us to use
the Sherman-Morrison formula, which reads
\begin{equation}
\left( \vec D + \vec u \vec v^\text T \right)^{-1} = \vec D^{-1} -
\frac{\vec D^{-1} \vec u \vec v^\text T \vec D^{-1}}
{1+ \vec v^\text T \vec D^{-1} \vec u},
\end{equation}
(note that, while $\vec u \vec v^\text T$ is an $n_\text p \times
n_\text p$ matrix, $\vec v^\text T \vec D^{-1} \vec u$ is a scalar). In
the present case $\vec D = \text{diag} \{ k^-_{1j}, k^-_{2j}, k^-_{3j},
\ldots\}$, while the two vectors are $\vec u = \{ k^+_{1j}, k^+_{2j},
k^+_{3j}, \ldots \}$ and $\vec v = \{ a_1, a_2, a_3, \ldots \}$. Using
the above definitions, together with $\vec b^j = c_j \vec u$, we find
the following equilibrium solution of the $j$-th block:
\begin{equation}
\widetilde{\vecs\theta^j} = \left(\vec M^j\right)^{-1} \vec b^j = 
c_j \ \left( \vec D + \vec u \vec v^\text T \right)^{-1} \vec u = 
\frac{c_j \vec D^{-1} \vec u}{1+ \vec v^\text T \vec D^{-1} \vec u}.
\label{again1}
\end{equation}
A simple calculation gives
\begin{equation}
\vec{v}^T \vec{D}^{-1} \vec{u} = \sum_{n=1}^{n_\text p} a_n K_{nj}
\qquad \text{and} \qquad
\vec{D}^{-1} \vec{u} = \{ K_{1j}, K_{2j}, K_{3j}, \ldots\},
\label{again2}
\end{equation}
where $K_{ij} \equiv k^+_{ij}/k^-_{ij}$ is the equilibrium
constant. Combining Eqs.~\eqref{again1} and \eqref{again2}, and recalling
that $\widetilde{\vecs{\theta}^j}_i = \widetilde\theta_{ij}$, we finally
obtain Eq.~\eqref{eq:depletion_eq}.

\subsection{Hybridization kinetics under depletion by a single sequence}

Equation~\eqref{eq:ode_mat} can be analytically solved when depletion
occurs due to a single probe. In this case we can write
\begin{equation}\label{eq:ode_probei}
\dod{\theta_{ij}}{t} = k_{ij}^+ \left( c_j - \sum_{n=1}^{n_\text p} 
a_n\theta_{nj}\right) - k_{ij}^- \theta_{ij}
\approx k_{ij}^+ \left( c_j - a_1\theta_{1j}\right)
- k_{ij}^- \theta_{ij},
\end{equation}
where we have assumed that $a_n \ll c_j$ for $n > 1$. This condition
can be experimentally realized through a proper design of the probes
and choice of target concentrations. Under this assumption, one has a
set of independent equations for $\theta_{1j}$ that can be easily solved
\begin{equation}\label{eq:sol_probe1}
\theta_{1j} = \frac{c_jK_{1j}}{1+a_1K_{1j}}
\left[1-e^{-(a_1k_{1j}^++k_{1j}^-)t} \right],
\end{equation}
which is monotonically growing in time and approaches the stationary
value $\widetilde{\theta}_{1j}=c_jK_{1j}/(1+a_1K_{1j})$. One can plug
Eq.~\eqref{eq:sol_probe1} in \eqref{eq:ode_probei} to solve for the
remaining $\theta_{ij}$ with $i > 1$. The result is
\begin{equation}\label{eq:sol_probei}
 \theta_{ij} = \frac{c_jK_{ij}}{1+a_1K_{1j}}\left[
            1 - e^{-t/\tau_{ij}} + \frac{k_{ij}^-}{k_{1j}^-} 
            \frac{a_1k_{1j}^+}{a_1k_{1j}^++k_{1j}^--k_{ij}^-}
            \left(e^{-t/\tau_{ij}}-e^{-t/\tau_{1j}}\right)
            \right],
\end{equation}
where $\tau_{ij} \equiv (a_1k_{ij}^+\delta_{i1} +  k_{ij}^-)^{-1}$
is a characteristic time. Note that by setting $i=1$ in
Eq.~\eqref{eq:sol_probei}, one recovers Eq.~\eqref{eq:sol_probe1},
as the third term within the square brackets vanishes. Thus,
Eq.~\eqref{eq:sol_probei} can be used as a general solution of the
problem for all $i$ and $j$.



\providecommand{\latin}[1]{#1}
\makeatletter
\providecommand{\doi}
  {\begingroup\let\do\@makeother\dospecials
  \catcode`\{=1 \catcode`\}=2 \doi@aux}
\providecommand{\doi@aux}[1]{\endgroup\texttt{#1}}
\makeatother
\providecommand*\mcitethebibliography{\thebibliography}
\csname @ifundefined\endcsname{endmcitethebibliography}
  {\let\endmcitethebibliography\endthebibliography}{}

\clearpage
\begin{center}
\edit{\sl{For Table of Contents only}}

\fedit{\includegraphics[width=0.5\columnwidth]{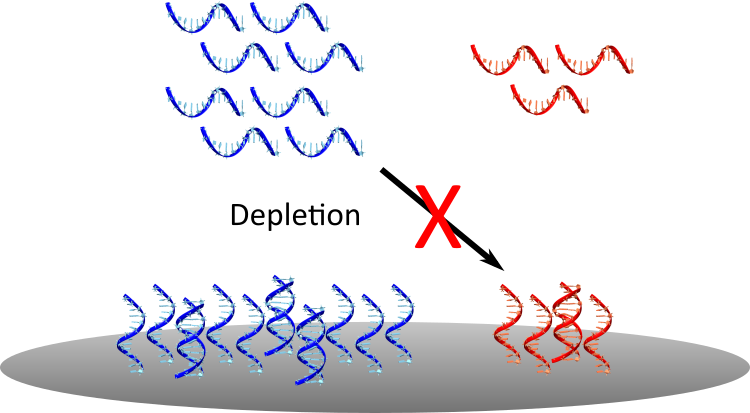}}

\edit{\sl{For Table of Contents only}}

\end{center}
\end{document}